\documentstyle[aps,epsf,twocolumn]{revtex}

\begin{document}
\draft
\title{Deceptive Apparent Nonadiabatic Magnetization Process}

\author{Keiji SAITO and Seiji MIYASHITA}

\address{
Department of Applied Physics, School of Engineering \\ 
University of Tokyo, Bunkyo-ku, Tokyo 113}

\author{Hans De RAEDT}

\address{Institute for Theoretical Physics and
Materials Science Centre,\\
University of Groningen, Nijenborgh 4,
NL-9747 AG Groningen, The Netherlands}

\date{\today}
\maketitle
\begin{abstract}
We discuss the effect of the thermal environment
on the low-temperature response of the magnetization of uniaxial magnets
to a time-dependent applied magnetic field.
At very low temperatures 
the steps-wise magnetization curves
observed in molecular magnets such as ${\rm Mn}_{12}$ and
${\rm Fe}_8$ display little temperature dependence
where the apparent thermal assisted process are suppressed. 
However the changes of the magnetization at each step 
cannot be analyzed directly in the view point of 
a quantum mechanical nonadiabatic transition.
In order to explain this deceptive apparent nonadiabatic behavior,
we study the quantum dynamics of the system weakly coupled to a thermal
environment and propose a relation between the observed 
magnetization steps and the quantum mechanical transition probability
due to the nonadiabatic transition. 
\end{abstract}
\noindent \\
Magnetization processes of nanoscale molecules such as Mn$_{12}$ and Fe$_{8}$
have attracted much interest.
For such small systems the discreteness
of energy level plays an important role and staircase structures of the 
response of the magnetization to a sweeping magnetic field
have been observed \cite{exp1,exp2,exp3,exp4,exp5,exp6,expP}.
The staircase is explained as a quantum mechanical transition
at the avoided level crossing points, where levels of the Hamiltonian 
become almost degenerate, and form repulsive structures as shown in
Fig.$1$, which has been called resonant tunneling.
This quantum mechanical transition has been studied from 
the point of view of the nonadiabatic transition
\cite{miya95,miya96,RMSGG97,MSD98}.
There are two characteristic features of each nonadiabatic 
transition \cite{RMSGG97}.
One is the localization of the transition because it occurs only around
avoided level crossing points. The other is the dependence of the transition 
probability on sweeping rate of the magnetic field, the energy gap and 
the gradients of the levels.
Since at each avoided level crossing point only two levels play an important 
role, the transition probability can be described by well-known the
Landau-Zener-St\"{u}ckelberg (LZS)  mechanism \cite{Landau,Zener,St}.

However the behavior of these magnetic systems can easily be affected
by thermal fluctuations even at low temperatures,
because the energy scales involved are rather small.
At relatively high temperatures ($T \sim 1$K)
the temperature dependence of the magnetization process is very significant,
where excitations to higher levels provide other channels of resonance 
tunneling which is called thermally assisted resonant tunneling 
\cite{GC97,FRVGS98,LBF98}. The external noise may affect the LZS mechanism 
itself which has been also studied \cite{KN98,DZ97,G97}.

On the other hand, at very low temperatures ($T \sim 60$ mK), the magnetization
curve shows very little change with temperatures and only 
quantum mechanical phenomena seem to be dominant \cite{expP}.
However, as we will show below, even at such low temperatures
thermal fluctuations cause inevitable effects which prevent
a direct application of mechanism of the nonadiabatic transition.

In this letter we investigate the effect of the thermal environment
at very low temperatures on nonadiabatic transitions and find a relation
between the observed data and the true quantum mechanical transition
probability, from which the energy gap at the avoided level crossing
point via the LZS formula can be deduced.

Let us consider the change of magnetization when the external field 
is swept from a negative value to a positive value. Initially
the system is assumed to be in the ground state with
the magnetization $m_0 \simeq -S$ (approximately). As the field increases,
the state with $m_0$ crosses states with the magnetization 
$S,$ $S-1,\cdots ,$ and $0$. At each avoided level crossing point 
a nonadiabatic transition occurs (Fig.$1$).
\begin{flushleft}
\epsfxsize=8cm \epsfysize=4.5cm \epsfbox{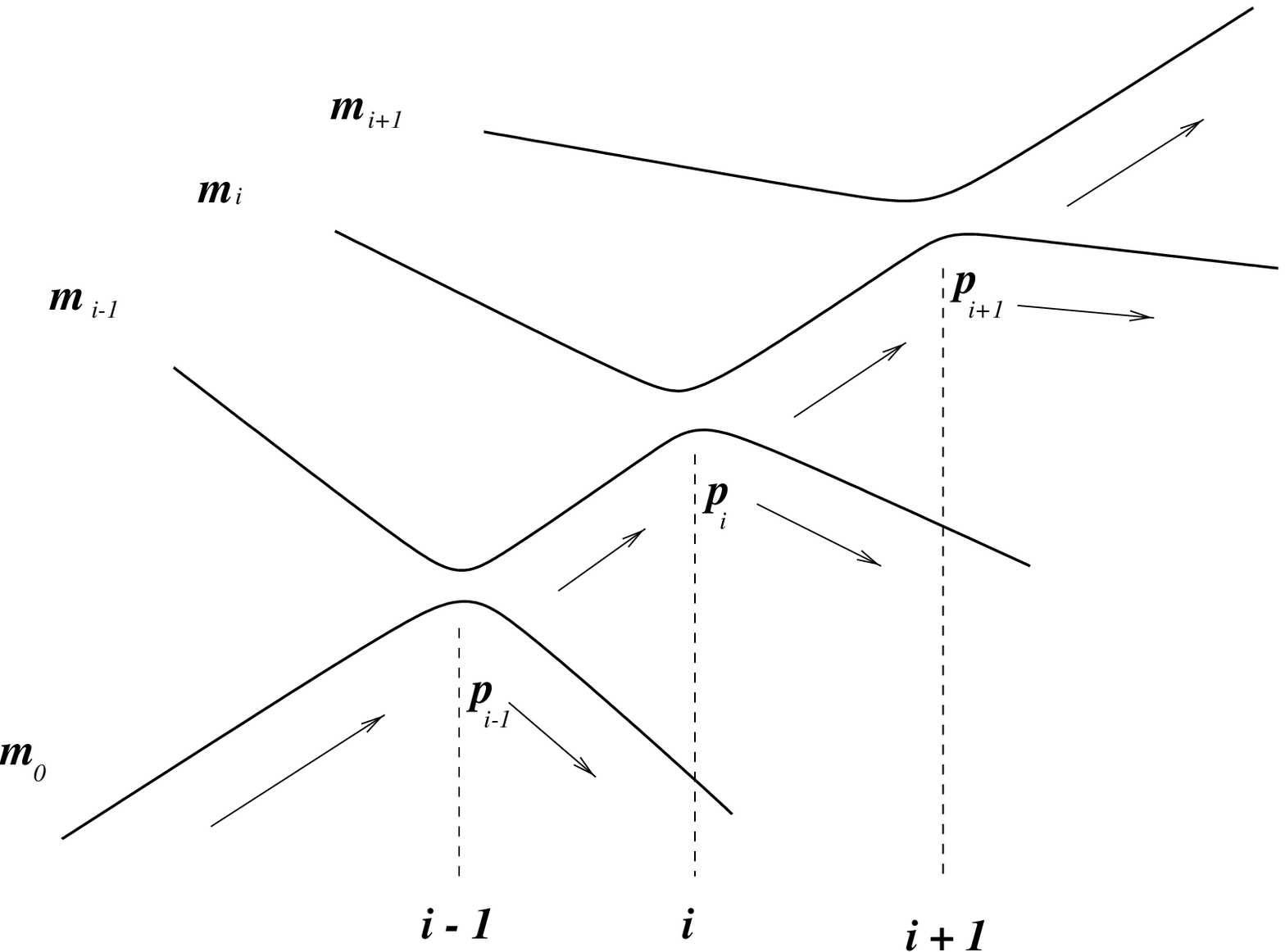} 
{\rm Fig.1 Schematic energy level diagram and 
the nonadiabatic transitions. $p_{i}$ denotes the probability that
the system remains in the same eigenstate. }
\end{flushleft}
We assign numbers $i$ $(i=1,2,\cdots )$ for the avoided level crossing point 
where the state of $m_0$ crosses a state with 
$m_i \simeq S-i+1$ ($= S, S-1,\cdots$, respectively). 
Let $p_{i} $ denote the probability staying the same
level at the $i$th avoided level crossing point.
For pure quantum dynamical case, we have the following relation between
the change of the observed magnetization at the crossing point $i$, 
$\Delta M_i \equiv M_{i}-M_{i-1}$ and the transition probabilities
$\{ p_i \}$:
\begin{eqnarray}
\Delta M_i &=& 
\prod_{n=1}^{i-1}(1-p_{n} ) 
\left\{ \left[ m_0 (1-p_{i}) + m_i  p_{i}\right] - m_0 \right\} ,
\label{magtom}
\end{eqnarray}
where $M_i$ is the observed magnetization between avoided level 
crossing point $i$ and $i+1$.
By this relation (\ref{magtom}), all the transition probabilities 
$\{ p_{i} \}$ are obtained from the magnetizations in pure quantum cases. 

In the experiment of Perenboom et al. for Mn$_{12} (S=10)$ 
($T = 59$mK) \cite{expP},
shape of the magnetization process seems to saturate with the lowering 
of the temperature. 
When we analyze the data using the relation (\ref{magtom}), 
we cannot find any consistent 
set of the transition probabilities $\{p_i\}$. In the experiment,
the steps-wise changes of the magnetization occur at the avoided level
crossing points where the state with $m_0 \simeq -10$ crosses with states 
with $m_i \simeq 3$, $2$, $1$ and $0$ ($i=7,8,9,$ and $10$, respectively). 
The changes of the magnetization at 
the points are $0.62$, $3.54$, $8.00$ and $6.77$, respectively.
The relation (\ref{magtom}) yields $p_7=0.0480$, $p_8=0.315$, $p_9=1.13$, 
and $p_{10}=-7.976$, in contradiction to the trivial condition 
$0\le p_i \le 1$. Therefore a naive application of nonadiabatic 
transition theory fails to explain the saturated 
magnetization curve in the very low temperature.

We attribute this failure to the effect of thermal environment
even at such a low temperature.
In terms of the potential picture (Fig.2),
the states with $M>0$ belong to the right valley and we expect that
these states easily relax to the bottom of the valley, i.e. to the state
with $M=S$.
Thus, once a quantum mechanical transition from the metastable state of $M=-S$
to a state of $M>0$ takes place,
the state is expected to relax easily
to the lowest level due to some dissipation mechanism in the absence of an 
energy barrier. In the case of pure quantum transition,
such a relaxation to the state of $M=S$ is prohibited because the levels 
of the states are separated far away.
If the time scale of dissipation is much shorter than 
that of the system and scale on which the magnetic field changes,
the transfer to the lowest state takes a short time. 
As a result the magnetization curve will show a staircase
as in the case of pure quantum dynamics, but
the change of magnetization at each step is different because
of the relaxation transition $M \rightarrow S$ instead of $S-i+1$.
We call this steps-wise magnetization process in a dissipative environment
``a deceptive apparent nonadiabatic magnetization process''.
\begin{center}
\noindent
\epsfxsize=6.5cm \epsfysize=4.8cm \epsfbox{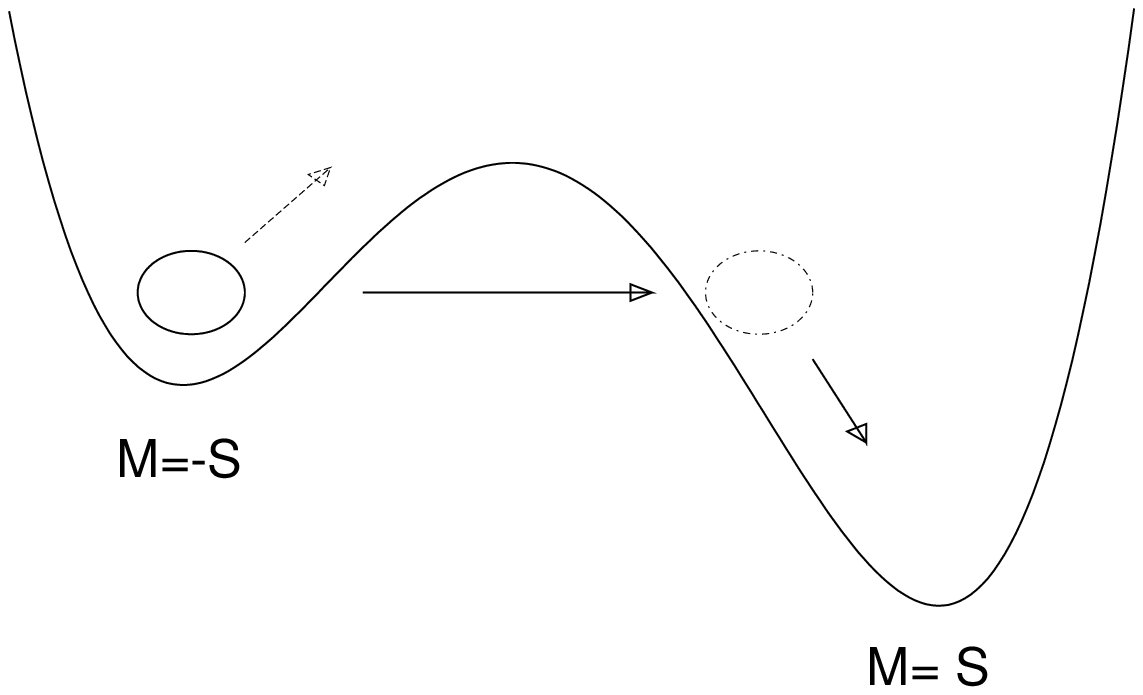} \\
{\rm Fig.2 Potential picture of the metastability.}
\end{center}

In this scenario, we assume the following three properties: (i) First quantum 
mechanical transition for $m_0 (\simeq -S)\rightarrow m_i$ occurs with the 
probability of the pure nonadiabatic (LZS) transition 
$\{ p_{i}^{\rm {LZS}}\}$, and then (ii) 
the relaxation from $m_i \rightarrow m_1 (\simeq S)$ 
occurs by some dissipation mechanism. (iii) There is no relaxation directly
from $m_0$ by the dissipation mechanism and thus the amount of magnetization 
change depends only on $\{ p_{i}^{\rm LZS}\}$ and does not depend on 
the temperature. 
Replacing $m_i $ by $m_1$ in the relation (\ref{magtom}), the change 
of the magnetization in this case is given by
\begin{eqnarray}
\Delta M_i &=& 
\prod_{n=1}^{i-1}(1-\tilde{p}_{n} ) 
\left\{ \left[ m_0 (1-\tilde{p}_{i}) + m_1 \tilde{p}_{i} \right]
- m_0 \right\} .
\label{magtomx}
\end{eqnarray}
Using the data of \cite{expP} now yields
a reasonable solution for the $\{\tilde p_i \}$'s:
$\tilde p_7=0.0313$, $\tilde p_8=0.185$, $\tilde p_9=0.515$, and 
$\tilde p_{10}=0.898$.

In order to demonstrate that the above three properties are really possible
at very low temperatures, we simulate a relaxation phenomena of a magnetic 
system which very weakly couples to the 
external bath. Here we use 
a quantum master equation \cite{STMPRE},
\begin{eqnarray}
\frac{\partial\rho(t)}{\partial t} &=& 
- i \left[{\cal H},\rho(t)\right] 
-\lambda
\left( \left[X,R\rho(t)\right] + \left[X,R\rho(t)\right]^{\dag} \right) , 
\label{CTTR}  
\end{eqnarray}
where 
\begin{eqnarray}
\langle k | R  | m \rangle &=& 
\zeta (\frac{E_{k} - E_{m}}{\hbar})
n_{\beta} ( E_{k} - E_{m} )  
\langle k | X  | m \rangle , \nonumber \\
\zeta (\omega ) &=& I(\omega ) -I(- \omega) ,
\quad {\rm and} \quad
n_{\beta}( \omega ) =  \frac{1}{e^{\beta\omega} -1 } . \nonumber 
\end{eqnarray}
Here $\beta$ is an inverse temperature of the reservoir $1/T$,
and we set $\hbar=1 $.
$| {k} \rangle $ and $| {m} \rangle $ are the eigenstates of
${\cal H}$ with the eigenenergies $E_{{k}}$ and $E_{{m}}$,
respectively. $I(\omega )$
is the spectral density of the boson bath.
We take here the infinite number of phonons with the Ohmic 
dissipation $I(\omega ) = I_{0} \omega $ \cite{GSI88}. As a more realistic
bath for the experimental situation at very low temperature,
we may take the dipole-field from the nuclear spins \cite{PS96} 
or other types of spectrum such as super-Ohmic type. 
$X$ is an operator of the magnetic system that interacts linearly with
the bosons of the reservoir.
The relaxation process can be affected by the form of interaction of
the system with the thermal bath, i.e. by
the choice of $X$. Here we take $X= \frac{1}{2} \left( S_{x} + S_{z} \right)$.
Generally $X=S_{x}$ is more efficient than $X=S_{z}$ for the relaxation.
A detailed comparison with other choices will be presented elsewhere.
The alternate choices of concrete form of thermal bath, however, do not
cause any significant qualitative change because the couplings
to the bath is very weak.

For Mn$_{12}$, detailed form of the Hamiltonian has been 
proposed \cite{KDHpre}. However the energy gap of the Mn$_{12}$
is too small to observe the phenomena within the available computation time.
Thus, here, we demonstrate the qualitative features of the dynamics, i.e. the 
three properties (i), (ii), and (iii). We believe that the key ingredients 
of the general qualitative feature are the existence of the 
avoided level crossing points and weak coupling to the external bath. 
For the realistic model with much small energy gap, the features observed 
here should be realized in a much longer time scale.
Thus we adopt 
a minimal model of a uniaxial $S=10$ spin system
with the two ingredients:
\begin{eqnarray}
{\cal H} = -D S_{z}^{2} + \Gamma S_{x} - H_{\rm ext} (t) S_{z} ,
\label{HAM}
\end{eqnarray}
with a linearly increasing external field, $H_{\rm ext} = c  t -  H_{0}$
where $c$ is the sweeping velocity.
The transverse field $\Gamma$ represents the terms causing quantum 
fluctuations. We choose $D =0.1$, $\Gamma = 0.5$ throughout this letter.

In order to see the difference of relaxations between the case with and 
without the potential barrier, we compare typical two cases: $(1)$
$H_{\rm ext} = 0.05$ and $(2)$ $H_{\rm ext} = 0.15$
and set the sweep velocity $c=0$.
As the initial state we take the second level, as indicated in Fig.$3$.
\begin{flushleft}
\epsfxsize=8.5cm \epsfysize=5.0cm \epsfbox{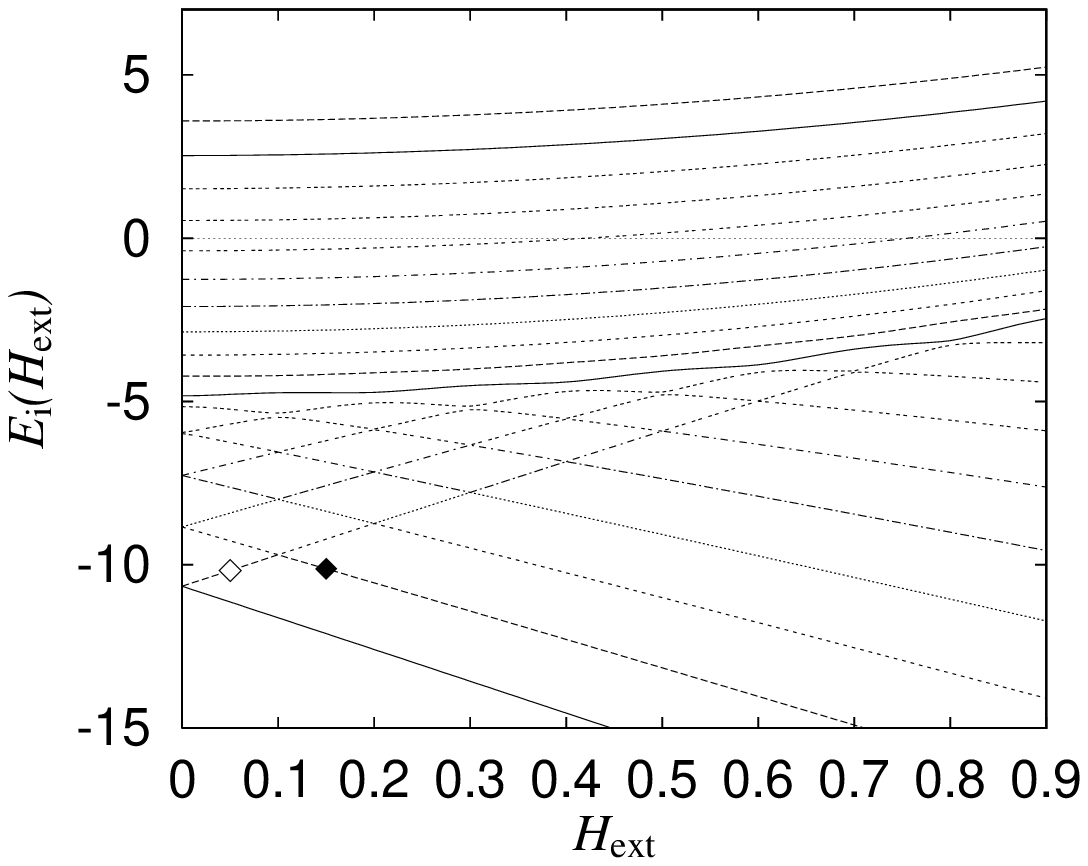} 
{\rm Fig.3  Energy level diagram of the model (\ref{HAM}) 
as a function of $H_{\rm ext}$.
The white and black diamonds correspond to the case $(1)$ and the 
case $(2)$, respectively.}
\end{flushleft}
The second level has $M\simeq -10$ in the case $(1)$ and $M\simeq 9$
in the case $2$. In the both cases, the ground state has $M\simeq 10$.
The parameters are set to $T = 0.1$, $I_{0} =1.0$, and 
$\lambda = 1.0 \times 10^{-4}$.
We study the relaxation for both cases by solving Eq.(\ref{CTTR}).
These probabilities are given by a diagonal element of $\rho (t)$, i.e.,
$\langle 1 | \rho (t) | 1 \rangle $ and $\langle 2 | \rho (t) | 2 \rangle$,
respectively. We observe almost no damping in the case $(1)$, whereas a
rather fast relaxation occurs in the case $(2)$. Thus at a fairly
low temperature, the thermal environment causes significant difference 
in the relaxation process depending on the presence of a potential barrier.
The difference between the cases $(1)$ and $(2)$ can be understood analyzing 
the matrix elements of Eq.(\ref{CTTR}).

We now investigate time evolution of the system for a sweeping field
$c = 1.0 \times 10^{-5}$ 
starting at $H_{0} = -0.05$. We study the case of pure quantum dynamics 
$(\lambda = 0)$[P] and the case with a weak dissipation
$(\lambda =1.0\times 10^{-4})$ [D]. These magnetization curves are shown 
in Fig.$4$. 
We show data for $H_{\rm ext} \ge 0.45$ because almost no change
is observed for $H_{\rm ext} < 0.5$.
\begin{flushleft}
\epsfxsize=8.0cm \epsfysize=6.0cm \epsfbox{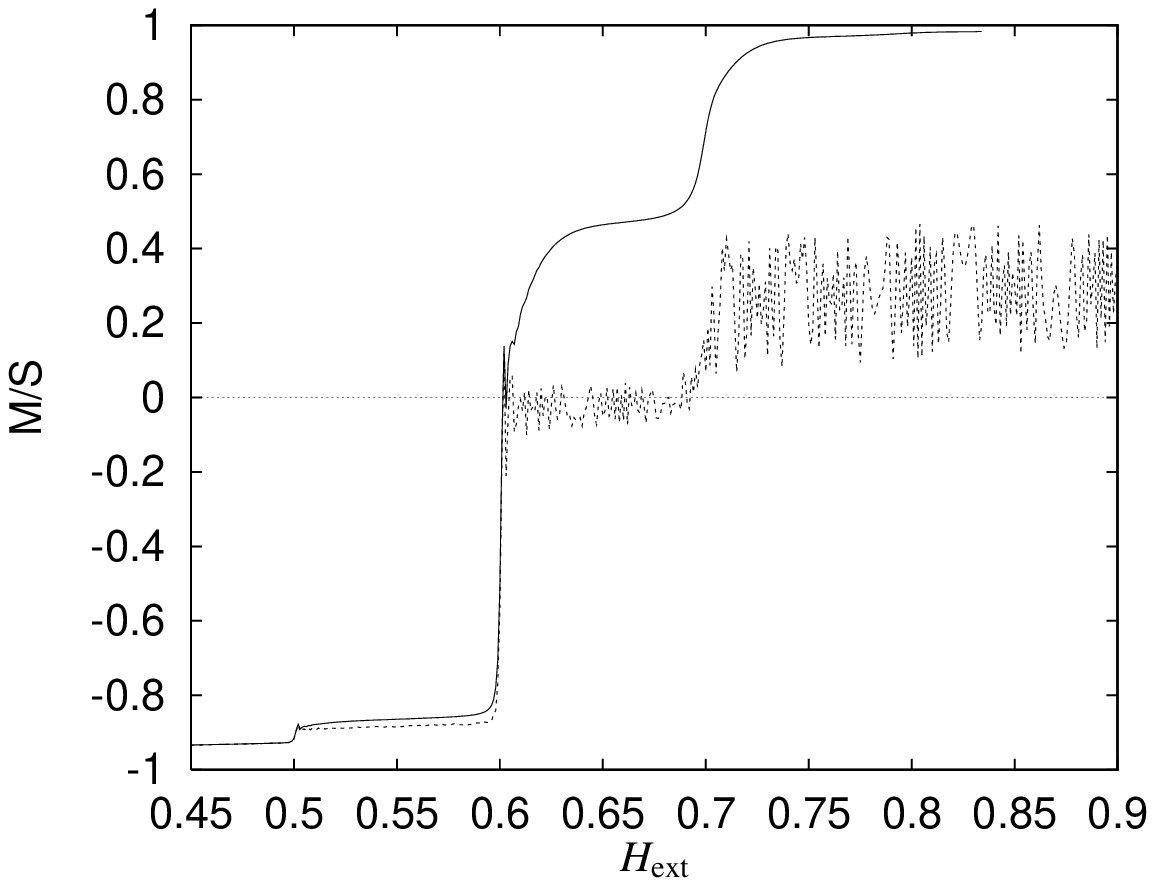} 
{\rm Fig.$4$ Magnetization as a function of $H_{\rm ext}$.
The dashed line denotes the pure quantum dynamics [P], and the
solid line denotes the dissipative quantum dynamics [D].}
\end{flushleft}
For the case [P], we observe oscillating behavior due to spin precession,
whereas in the case [D] this detailed structure is smoothed out by 
the dissipation.
We find steps-wise magnetization curves in both cases.
The changes of the magnetization are listed in Table I.
\begin{center}
\begin{tabular}{|c|c|c|}
\hline
Cross point $(m_0, m_i)$& \quad$\Delta M_{[{\rm P}]}$\quad 
                     & \quad$\Delta M_{[{\rm D}]}$\quad \\
\hline
(-10,5)&0.511&0.693\\
\hline
(-10,4)&8.32&13.3\\
\hline
(-10,3)&3.50&5.03\\
\hline
\end{tabular}
\end{center}
\begin{center}
 Table I The changes of magnetization. $\Delta M_{[{\rm P}]}$ and 
$\Delta M_{[{\rm P}]}$ are the changes for the case [P] and [D], respectively.
\end{center}
\begin{center}
\begin{tabular}{|c|c|c|c|c|c|c|c|}
\hline
           \,\, $i$ \qquad 
         &   $(m_0, m_i )$ \qquad
         & \quad${ p }_{[{\rm R}],i}$
         & \quad$p_{[{\rm P}],i}$\qquad 
         & \quad$p_{[{\rm D}],i}$\qquad 
         & \quad${\tilde p }_{[{\rm D}],i}$
         & \quad${\bar p }_{[{\rm D}],i}$
\qquad \\
\hline
$6$ & (-10,5) &0.0280 & 0.0341 & 0.0460 & 0.0346  & 0.0291 \\
\hline
$7$ & (-10,4) &0.730  & 0.616 & 0.995   & 0.688  & 0.716 \\
\hline
$8$ & (-10,3) &1.000 & 0.726  & 78.0 &  0.835  & 0.970  \\
\hline
\end{tabular}
\end{center}
\begin{center}
Table II The transition probabilities obtained by various ways, see the text.
\end{center}

From these data we estimate
the transition probabilities by the relation (\ref{magtom}) and 
(\ref{magtomx}) which are listed in Table II.
First we obtain the transition probabilities 
from the data in Table I setting $m_0 = -S$ and $m_{i} =S-i+1$.
From the data $\Delta M_{[D]}$, unacceptable probabilities 
$\{ p_{[{\rm D}],i}\}$ are deduced by 
the relation (\ref{magtom}), while acceptable ones 
$\{ \tilde{p}_{[{\rm D}],i} \}$ are obtained by the relation (\ref{magtomx}).
$\{ \tilde{p}_{[{\rm D}],i} \}$ agree with $\{ p_{[{\rm P}],i} \}$ 
obtained by the relation (\ref{magtom}) from the data $\Delta M_{[P]}$.
This agreement shows the three properties (i), (ii), and (iii) 
really realized in the present 
model and thus we can estimate the quantum mechanical nonadiabatic
transition by the relation (\ref{magtomx}).
Although the magnetization $m_i$ is almost constant: $m_0 \simeq -S, 
m_i \simeq S-i+1 \, (i\ge 1)$, they show a little dependence on the 
magnetic field $H_{\rm ext }$.
Taking the $H_{\rm ext}$ dependence of $m_{i}$ into consideration, 
we also calculated the transition probabilities in the case [D] 
with (\ref{magtomx}). They are shown as $\{ \bar{p}_{[{\rm D}],i}\}$.
We confirmed that $\{ \bar{p}_{[{\rm D}],i}\}$ agree 
with the probabilities 
$\{ p_{[{\rm R}],i}\}$ directly obtained from 
the diagonal elements of the density matrix.
The difference between $\tilde{p}_{[{\rm D}],i}$ and 
$\bar{p}_{[{\rm D}],i}$ simply come from the large value of $\Gamma$ 
for the convenience of simulation. If $\Gamma$ is very small as the 
case of the experiment, $m_i$ is very close to $S-i+1$ and it is expected 
that $\tilde{p}_{[{\rm D}],i}$ and $\bar{p}_{[{\rm D}],i}$ are very close.
We present the time evolution of $\langle i | \rho | i \rangle$
in Fig.$5$. This figure explicitly shows the three properties (i), (ii), and 
(iii). 
\begin{center}
\epsfxsize=8.0cm \epsfysize=6.0cm \epsfbox{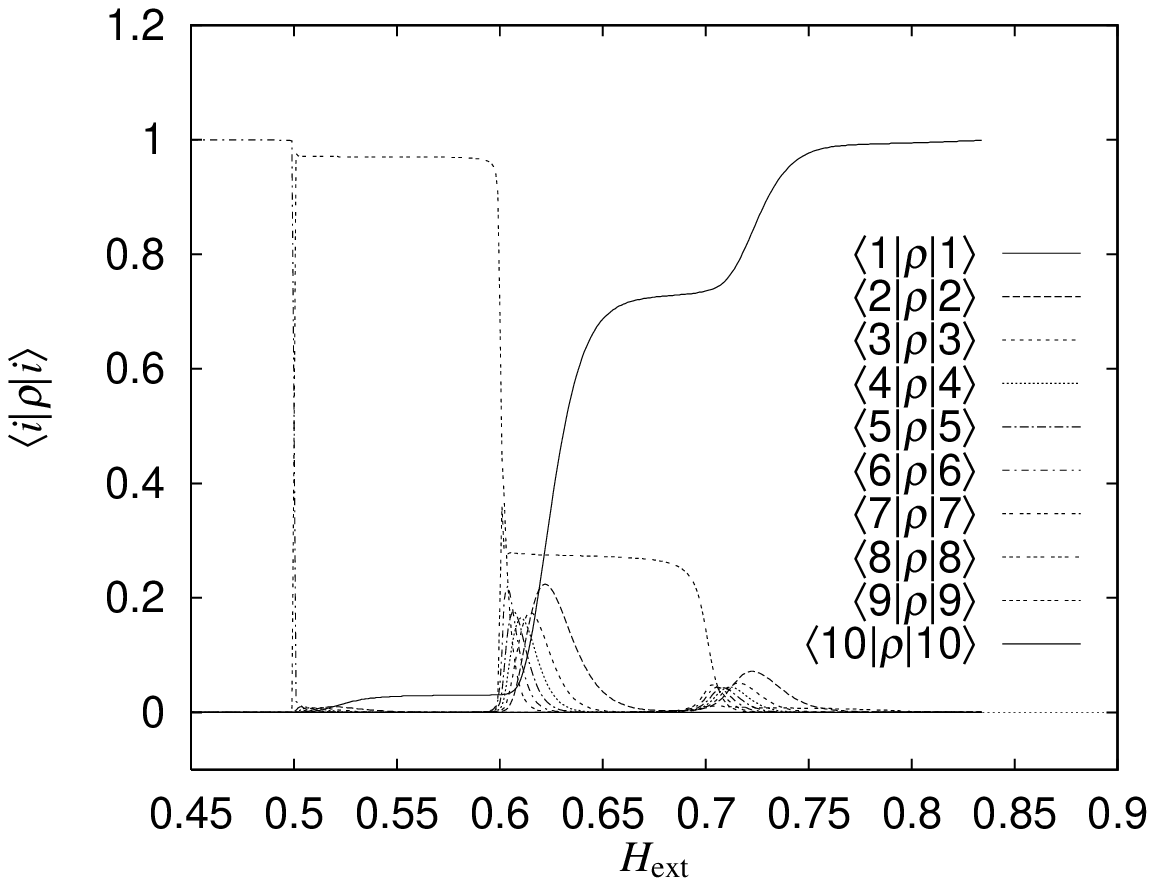} 
{\rm Fig.5 The time evolution of the probability of individual levels.}
\end{center}

We estimate the energy gap from the transition probabilities 
with the extended LZS formula $p_{i}^{\rm LZS}$:  
\begin{equation}
p_{i}^{\rm LZS} = 
1-\exp\left(-\frac{ \pi (\Delta E_{i})^2}{2 (m_i - m_0 ) c}\right) ,
\label{ELZS}
\end{equation}
where $c$ is the changing rate of the Zeeman energy.
Using $\{ \tilde{p}_{[{\rm D}],i} \}$, we obtain the energy gaps for 
the avoided level crossings as 
$\Delta E_6 = 1.83 \times 10^{-3},$  
$\Delta E_7 = 10.1 \times 10^{-3}$.
These estimates agree with the correct value 
$\Delta E_6 = 1.54 \times 10^{-3}$ and 
$\Delta E_7 = 10.0 \times 10^{-3}$ directly obtained from 
the energy levels \cite{E8}.
If we use $\bar{p}_{[{\rm D}],i}$, we have, of course, almost 
complete values, $\Delta E_6 = 1.57 \times 10^{-3}$, 
$\Delta E_7 = 9.9 \times 10^{-3}$.
Thus we conclude that we can estimate the energy gap from the deceptive 
apparent magnetization by the relation (\ref{magtomx}).

In summary, we have considered a mechanism for
deceptive apparent nonadiabatic magnetization process
which is relevant when the temperature is very low and no temperature
dependence is observed apparently and propose the general relation 
(\ref{magtomx}) between the steps in the magnetization and the 
energy-level splittings at very low temperature.
With the relation we have estimated the quantum transition rate 
$\{ p_i \}$
in the low temperatures \cite{expP}. 
We demonstrated an example of deceptive apparent nonadiabatic 
magnetization process in a minimal model with the 
avoided level crossing points and weak coupling to the external bath. 
Elsewhere we will report on our investigation of
the energy gaps $\{ \Delta E_{i}\}$ of Mn$_{12}$ and Fe$_{8}$ based on the
detailed information of the values of jumps and the scanning speed $c$.

The present study is partially supported by the Grant-in Aid for
Research from the Ministry of Education, Science and Culture.


\begin{references}
\bibitem{exp1} J. R. Friedman and M. P. Sarachik, T. Tejada and
R. Ziolo, Phys. Rev. Lett. {\bf 76}, 3830 (1996).
\bibitem{exp2} L. Thomas, F. Lionti, R. Ballou, D. Gatteschi,
R. Sessoli and B. Barbara, Nature {\bf 383}, 145 (1996).
\bibitem{exp3} J. M. Hernandez, X. X. Zhang, F. Luis, and T. Tejada,
J. R. Friedman, M. P. Sarachik and R. Ziolo,
Phys. Rev. B {\bf 55}, 5858 (1997).
\bibitem{exp4} L. Thomas et al., Nature {\bf 383}, 145 (1996).
\bibitem{exp5} F. Lionti, L. Thomas, R. Ballou, Barbara, A. Sulpice,
R. Sessoli and Gatteschi, J. Appl. Phys. {\bf 81}, 4608 (1997).
\bibitem{exp6} C. Sangregorio, T. Ohm, C. Paulsen, R. Sessoli, D. Gatteschi, 
Phys. Rev. Lett. {\bf 78}, 4645 (1997).
\bibitem{expP} J. A. A. J. Perenboom, J. S. Brooks, and S. Hill, and T. Hathaway, 
and N. S. Dalal, Phys. Rev. {\bf B} 330 (1998).
\bibitem{miya95} S. Miyashita, J. Phys. Soc. Jpn. {\bf 64}, 3207 (1995).
\bibitem{miya96} S. Miyashita, J. Phys. Soc. Jpn. {\bf 65}, 2734 (1996).
\bibitem{RMSGG97}
H. De Raedt, S. Miyashita, K. Saito, D. Garc\'{i}a-Pablos, and N. Garc\'{i}a,
Phys. Rev. B, {\bf 56} 11761 (1997).
\bibitem{MSD98}
S. Miyashita, K. Saito, and H. De Raedt, Phys. Rev. Lett. {\bf 80},
1525 (1998).
\bibitem{Landau} L. Landau, Phys. Z. Sowjetunion {\bf 2}, 46 (1932).
\bibitem{Zener} C. Zener, Proc. R. Soc. London, Ser. A{\bf 137}, 696 (1932).
\bibitem{St} E. C. G. St\"uckelberg, Helv. Phys. Acta {\bf 5}, 369 (1932).
\bibitem{GC97}
D. A. Granin and E. M. Chudnovsky, Phys. Rev. B {\bf 56}, 11102 (1997)
\bibitem{FRVGS98}
A. Fort, A. Rettori, J. Villain, D. Gatteschi, and R. Sessoli, 
Phys. Rev. Lett. {\bf 80}, 612 (1998) 
\bibitem{LBF98}
F. Luis, J. Bartolom\'{e}, and F. Fern\'{a}ndez, Phys. Rev. B {\bf 57}, 
505 (1998)
\bibitem{KN98}
Y. Kayanuma and H. Nakayama, Phys. Rev. B {\bf 57}, 13099 (1998)
\bibitem{DZ97}
V. V. Dobrovitski and A. K. Zvezdin, Europhys. Lett. {\bf 38}, 377 (1997)
\bibitem{G97}
L. Gunther, Europhys. Lett. {\bf 39}, 1 (1997) 
\bibitem{STMPRE}
K. Saito, S. Takesue, S. Miyashita, unpublished, cond-mat/9810069.
\bibitem{GSI88}
H, Grabert, P. Schramn, and G. Ingold, 
Phys.\  Rep.\  {\bf 3}, 115 (1988).
\bibitem{PS96}
 N. V. Prokof'ev and P. C. E. Stamp, J. Low Temp. Phys. 104, 143 (1996).
\bibitem{KDHpre}
M. I. Katsnelson, V. V. Dobrovitski and B. N. Harmon, cond-mat/9807176.  
\bibitem{E8}
Due to the large value of $\Gamma$ and thus $p_8 \sim 1$,
it is difficult to estimate precisely the concrete value of $\Delta E_8$.
\end{references}
\end{document}